\documentclass[%
reprint,
superscriptaddress,
amsmath,amssymb,
aps,
]{revtex4-2}

\usepackage{graphicx}
\usepackage{dcolumn}
\usepackage{bm}
\usepackage{hyperref}

\usepackage{txfonts}
\usepackage{braket}
\usepackage{comment}
\usepackage{color}

\begin{document}

\newcommand\red[1]{\textcolor{red}{#1}}
\newcommand\blue[1]{\textcolor{blue}{#1}}

\title{Universality of active and passive phase separation in a lattice model}

\author{Kyosuke Adachi}
\affiliation{Nonequilibrium Physics of Living Matter RIKEN Hakubi Research Team, RIKEN Center for Biosystems Dynamics Research, 2-2-3 Minatojima-minamimachi, Chuo-ku, Kobe 650-0047, Japan}
\affiliation{RIKEN Interdisciplinary Theoretical and Mathematical Sciences Program, 2-1 Hirosawa, Wako 351-0198, Japan}

\author{Kyogo Kawaguchi}
\affiliation{Nonequilibrium Physics of Living Matter RIKEN Hakubi Research Team, RIKEN Center for Biosystems Dynamics Research, 2-2-3 Minatojima-minamimachi, Chuo-ku, Kobe 650-0047, Japan}
\affiliation{RIKEN Cluster for Pioneering Research, 2-2-3 Minatojima-minamimachi, Chuo-ku, Kobe 650-0047, Japan}
\affiliation{Universal Biology Institute, The University of Tokyo, Bunkyo-ku, Tokyo 113-0033, Japan}

\date{\today}

\begin{abstract}
The motility-induced phase separation (MIPS) is the spontaneous aggregation of active particles, while equilibrium phase separation (EPS) is thermodynamically driven by attractive interactions between passive particles.
Despite such difference in the microscopic mechanism, similarities between MIPS and EPS like free energy structure and critical phenomena have been discussed.
Here we introduce and analyze a 2D lattice gas model that undergoes both MIPS and EPS by tuning activity and interaction parameters.
Based on simulations and mean-field theory, we find that the MIPS and EPS critical points are connected through a line of nonequilibrium critical points.
According to the size scaling of physical quantities and time evolution of the domain size, both the static and dynamical critical exponents seem consistent with the 2D spin-exchange Ising universality over the whole critical line.
The results suggest that activity effectively enhances attractive interactions between particles and leaves intact the critical properties of phase separation.
\end{abstract}

\maketitle

\textit{Introduction.}
In active matter systems, each element converts external energy into self-propulsion, which can lead to unique nonequilibrium phase transitions like flocking~\cite{Vicsek1995,Gregoire2004,Solon2013,Solon2015a,Solon2015b,Martin2020}, active nematic ordering~\cite{Nishiguchi2017,Kawaguchi2017,Duclos2020}, and microphase separation~\cite{Tjhung2018,Caporusso2020,Shi2020}.
In particular, the motility-induced phase separation (MIPS)~\cite{Cates2015} is a representative activity-induced phase transition found in simulation studies~\cite{Tailleur2008,Thompson2011,Fily2012} and observed both in biological~\cite{Liu2019,Fragkopoulos2020} and artificial~\cite{Buttinoni2013} systems.
MIPS represents the aggregation of self-propelled particles with crowding/repulsive interactions~\cite{Cates2015}, markedly different from equilibrium phase separation (EPS), which is thermodynamically driven by attractive interactions between passive particles.
Despite such differences in the microscopic mechanism, similarities between MIPS and EPS have been discussed~\cite{Tailleur2008}, and recently a generalized free energy functional for MIPS has been proposed~\cite{Solon2018a,Solon2018b} and applied to microscopic models~\cite{Kourbane2018}.

It is interesting to consider how the concepts of critical phenomena and universality~\cite{Hohenberg1977} can be applied to active matter systems~\cite{Chen2015,Chen2020}.
According to numerical studies of active lattice gas models~\cite{Partridge2019} and Active Ornstein-Uhlenbeck particles~\cite{Maggi2020}, the MIPS critical point in two dimensions seems to belong to the 2D Ising universality class, which is the same as for the EPS critical point.
Theoretically, the perturbative renormalization group (RG) analysis of the Active Model B+ has shown that weak activity does not change the universality class of phase separation~\cite{Caballero2018}.
On the other hand, the critical exponents of MIPS observed in simulations of Active Brownian particles have been incompatible with the Ising universality~\cite{Siebert2018,Kolb2020}.
Additionally, in simulations of Active Brownian particles with attractive interactions, phase separation is stabilized for weak or strong activity but suppressed for moderate activity~\cite{Redner2013}, suggesting that activity can also effectively suppress the attractive interaction.
Thus, it is still unclear if there exists a microscopic model that shows MIPS and EPS with the same Ising universality.

To clarify the relation between the MIPS and EPS critical points, it is natural to ask if we can find a critical line which connects them by tuning parameters of a microscopic model~\cite{Paoluzzi2016,Paoluzzi2020}.
If the critical line exists, the next question is whether the whole line, which corresponds to nonequilibrium critical points for any nonzero activity, belongs to the Ising universality class.
In this Letter, we address these questions by constructing and analyzing a lattice gas model with both activity and attractive interactions, which undergoes both MIPS and EPS.
First, based on numerical simulations and mean-field theory, we find that the MIPS and EPS critical points are indeed connected through a critical line.
Then, using the finite-size scaling analysis and examining time evolution, we conclude that the whole critical line belongs to the 2D Ising universality class, which suggests that activity-induced violation of detailed balance is irrelevant for critical properties of phase separation.

\begin{figure*}[t]
\centering
\includegraphics[scale=0.8]{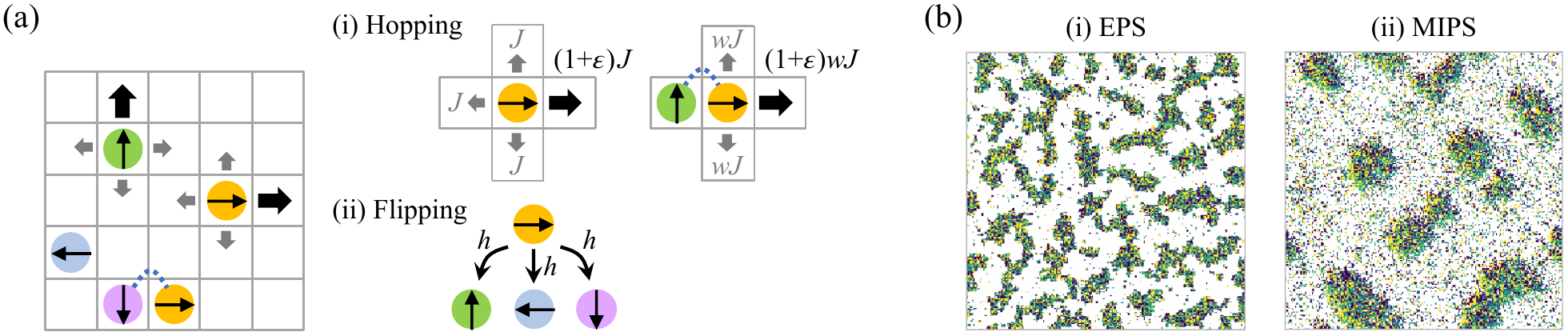}
\caption{(a) Lattice gas model with activity, nearest-neighbor interaction, and on-site exclusion.
Each particle with a spin can stochastically (i) hop to a nearest-neighbor site with a larger rate in the spin direction or (ii) flip the spin.
(b) Typical configurations of growing (i) EPS and (ii) MIPS in a square system ($L_x = L_y = 200$) with periodic boundary conditions.
The yellow, green, blue, and purple dots represent the particles with $s = \hat{x}$, $\hat{y}$, $-\hat{x}$, and $-\hat{y}$, respectively.
We used $\varepsilon = 0$, $U = -2$, $h / J = 0.01$, and $\overline{\rho} = 0.4$ with $5 \times 10^4$ MC steps for (i); $\varepsilon = 2$, $U = 0$, $h / J = 0.01$, and $\overline{\rho} = 0.4$ with $5 \times 10^3$ MC steps for (ii)~\cite{SM}.}
\label{Fig:Model}
\end{figure*}

\begin{figure*}[t]
\centering
\includegraphics[scale=0.8]{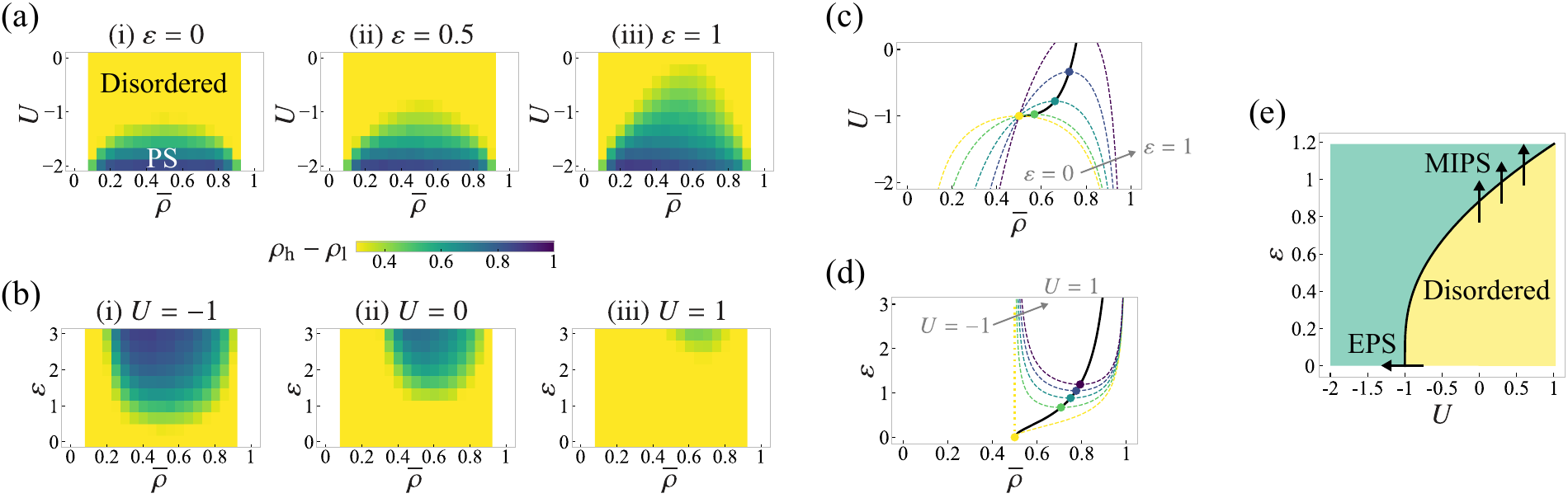}
\caption{(a, b) Numerically obtained phase diagrams in the (a) $\overline{\rho}$-$U$ and (b) $\overline{\rho}$-$\varepsilon$ planes for a rectangular system ($L_x = 40$ and $L_y = 4$).
The heatmap shows the density difference ($\rho_\mathrm{h} - \rho_\mathrm{l}$) between the high-density and low-density phases.
We took $480$ samples with $10^6$ or $10^5$ MC steps for (a) or (b), respectively~\cite{SM}.
(c, d) Mean-field critical points (circles) and spinodal lines (dashed lines) in the (c) $\overline{\rho}$-$U$ plane with $\varepsilon = 0, 0.25, 0.5, 0.75, 1$ and (d) $\overline{\rho}$-$\varepsilon$ plane with $U = -1, -0.5, 0, 0.5, 1$.
In (c) and (d), we also show the critical line (black line) projected in each plane.
(e) The mean-field critical line in the $U$-$\varepsilon$ plane with $\overline{\rho}$ satisfying Eq.~\eqref{Eq:A3}.
For $\varepsilon = 0$, the critical EPS transition occurs as we increase the attractive interaction (negative $U$); for $U \geq 0$, the critical MIPS transition occurs as we increase the activity ($\varepsilon$).
For all figures, we used $h / J = 0.01$.}
\label{Fig:PhaseDiagram}
\end{figure*}

\begin{figure*}[t]
\centering
\includegraphics[scale=0.8]{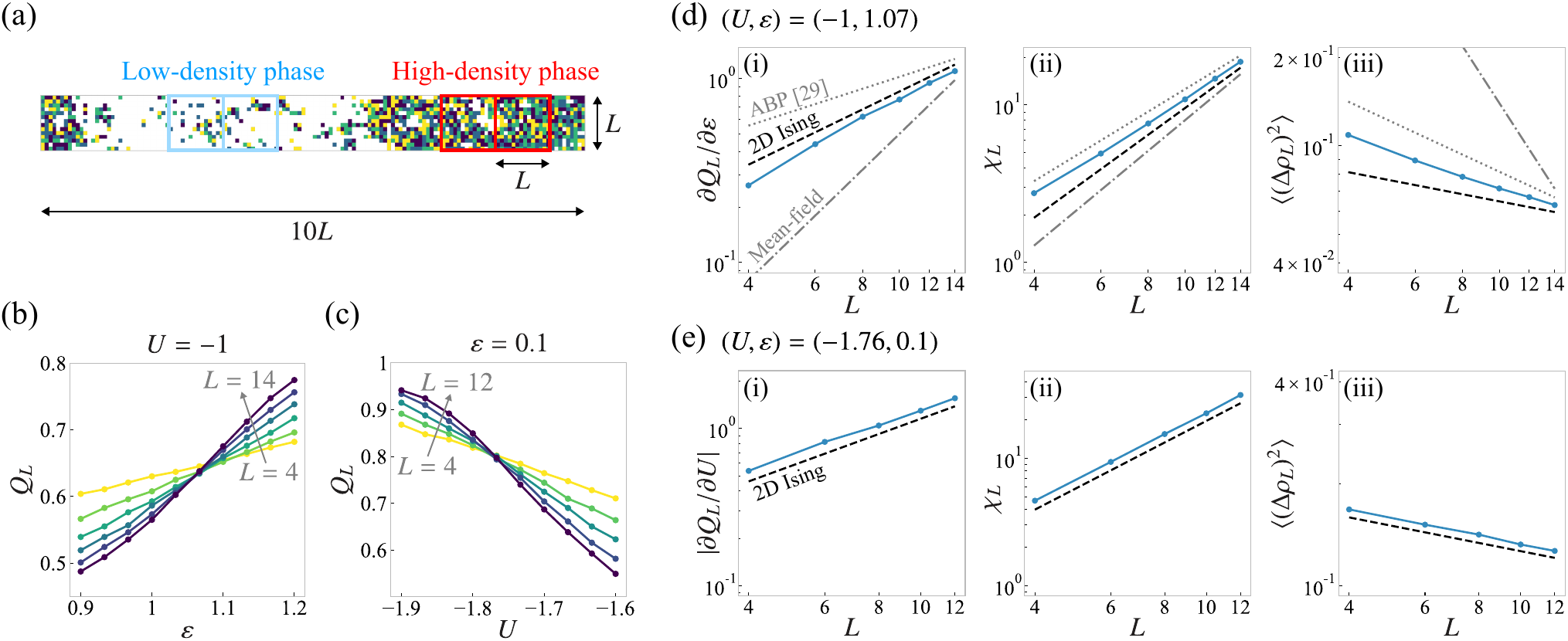}
\caption{(a) An example of configuration and four sub-boxes.
The $x$-coordinate of the left edge of each sub-box is $x_\mathrm{c} - L$, $x_\mathrm{c}$, $x_\mathrm{c} + 4 L$, and $x_\mathrm{c} + 5 L$ ($\mathrm{mod} \, 10 L$), where $x_\mathrm{c}$ is the center-of-mass $x$-coordinate.
(b, c) The Binder ratio as a function of a parameter for several sub-box sizes [(b) varying $\varepsilon$ with $L = 4, 6, 8, 10, 12, 14$ or (c) varying $U$ with $L = 4, 6, 8, 10, 12$].
(d, e) Size scaling of the (i) derivative of the Binder ratio ($\sim L^{1 / \nu}$), (ii) susceptibility ($\sim L^{\gamma / \nu}$), and (iii) density fluctuation ($\sim L^{-\beta / \nu}$).
For comparison, we show the size scalings for the 2D Ising model ($\nu = 1$, $\beta = 1 / 8$, and $\gamma = 7 / 4$; black dashed line), the Active Brownian particles (ABP)~\cite{Siebert2018} ($\nu = 1.5$, $\beta = 0.45$, and $\gamma = 2.2$; gray dotted line), and the mean-field Ising model ($\nu = 1 / 2$, $\beta = 1 / 2$, and $\gamma = 1$; gray dashed-dotted line).
For (b) and (d) [(c) and (e)], we performed 480 independent simulations and sampled 41 (21) configurations at intervals of $10^5$ MC steps  after $2 \times 10^6$ ($6 \times 10^6$) MC steps in each simulation with the random (fully phase-separated) initial configurations~\cite{SM}.
For all figures, we used $h / J = 0.01$ and $\overline{\rho} = 0.5$.
}
\label{Fig:FSS}
\end{figure*}

\begin{figure}[t]
\centering
\includegraphics[scale=0.8]{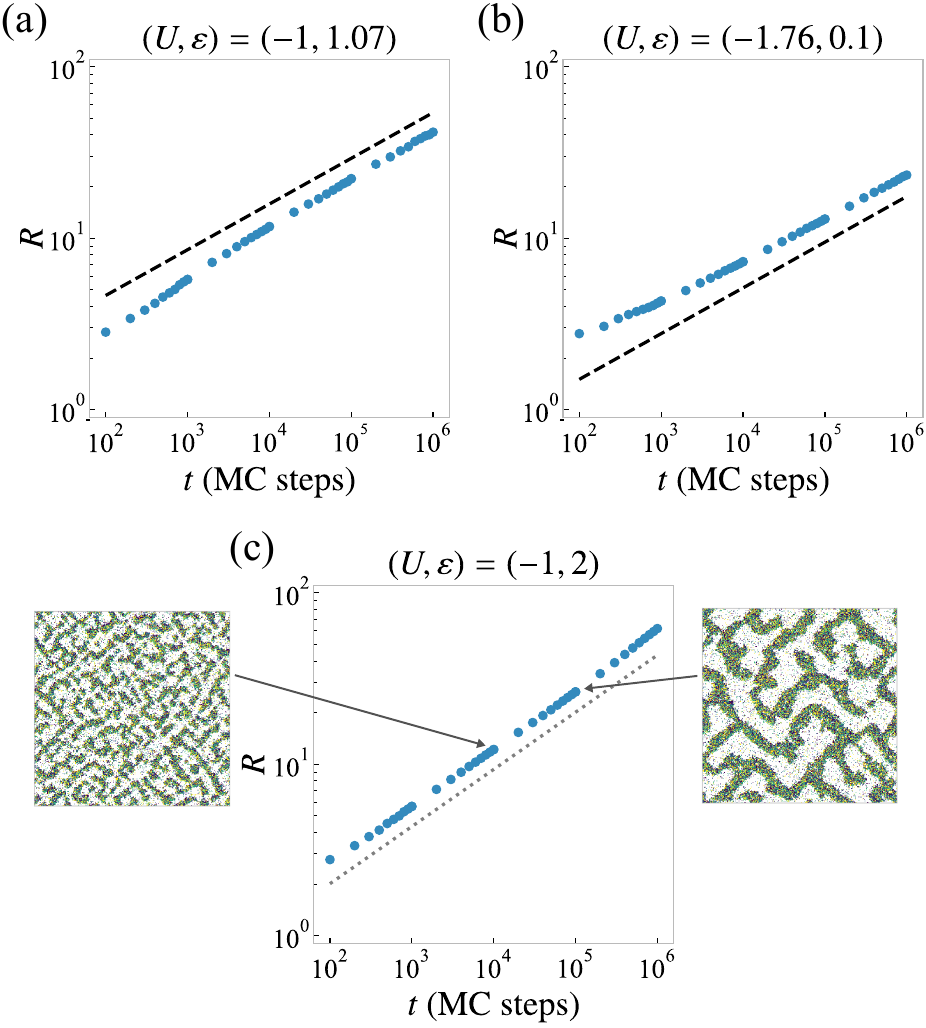}
\caption{Time evolution of the domain size $R (t) \sim t^{1/z}$ for two critical points [(a) $(U, \varepsilon) = (-1, 1.07)$ and (b) $(-1.76, 0.1)$] and an off-critical point [(c) $(-1, 2)$].
For comparison, we show the time scalings for the 2D spin-exchange Ising universality ($z = 15/4$; black dashed line) and the LSW law ($z = 3$; gray dotted line).
In (c), we show typical configurations for $10^4$ and $10^5$ MC steps.
For all figures, we performed 50 independent simulations for the $500 \times 500$ square lattice with $h/J = 0.01$ and $\overline{\rho} = 0.5$.}
\label{Fig:Dynamics}
\end{figure}

\textit{Model.}
To discuss both MIPS and EPS within a single framework, we consider a lattice gas model with both activity and nearest-neighbor interaction [Fig.~\ref{Fig:Model}(a)].
In this model, each particle with a spin $s$ ($= \hat{x}$, $\hat{y}$, $-\hat{x}$, or $-\hat{y}$) can stochastically (i) hop to a nearest-neighbor site if empty or (ii) flip the spin with a rate $h$, where $\hat{a}$ is the unit translation parallel to the $a$\nobreakdash-axis.
For hopping from site $i$ to an adjacent site $j$, we set a higher rate $(1 + \varepsilon) w_{i \to j} J$ if the hopping is in the same direction as the spin, and a lower rate $w_{i \to j} J$ otherwise, using the activity parameter $\varepsilon$ ($\geq 0$).
We set $w_{i \to j} = 1 - \tanh (\Delta E_{i \to j} / 2)$ with $k_\mathrm{B} T = 1$, where $\Delta E_{i \to j}$ is the increase in the total interaction energy due to hopping, with the nearest-neighbor interaction energy $U$ (repulsive for $U > 0$ and attractive for $U < 0$).
Note that the equilibrium heat-bath dynamics~\cite{Stoll1973,Binder1975,Binder1977} recovers for $\varepsilon = 0$. 
Following previous studies~\cite{Siebert2018,Partridge2019,Maggi2020}, we refer to the phase separation that occurs under $U \geq 0$ (with no attractive interactions) as MIPS.

As expected, EPS occurs for large negative $U$ [Fig.~\ref{Fig:Model}(b)(i)] in the case with $\varepsilon = 0$, whereas in the case with $U \geq 0$, MIPS occurs for large $\varepsilon$ [Fig.~\ref{Fig:Model}(b)(ii)].
The effective parameters in this model are $\varepsilon$, $U$, $h / J$, and the average density $\overline{\rho}$ ($0 < \overline{\rho} < 1$).
In the following Monte Carlo (MC) simulations~\cite{SM}, we set $h / J = 0.01$.
To reduce the interface effects and apply the sub-box method in the finite-size scaling analysis, we consider rectangular systems with an aspect ratio of 10:1, except when measuring the dynamical critical exponent.

\textit{Connection between MIPS and EPS critical points.}
In Fig.~\ref{Fig:PhaseDiagram}, we show the steady-state phase diagrams in the (a) $\overline{\rho}$-$U$ and (b) $\overline{\rho}$-$\varepsilon$ planes.
The heatmap represents the density difference between the high-density and low-density phases ($\rho_\mathrm{h} - \rho_\mathrm{l}$), which is the order parameter for phase separation.
Note that Fig.~\ref{Fig:PhaseDiagram}(a)(i) is the phase diagram for EPS since $\varepsilon = 0$, and Figs.~\ref{Fig:PhaseDiagram}(b)(ii) and (iii) are the phase diagrams for MIPS since $U \geq 0$.
From Fig.~\ref{Fig:PhaseDiagram}(a) [Fig.~\ref{Fig:PhaseDiagram}(b)], we find that the critical point, located at the tip of the phase boundary in the $\overline{\rho}$-$U$ ($\overline{\rho}$-$\varepsilon$) plane, moves continuously as we change $\varepsilon$ ($U$).
Consequently, in the $\overline{\rho}$-$U$-$\varepsilon$ space, there is a critical line which connects the EPS and MIPS critical points.

In the following, we consider the qualitative behavior of the critical line by a mean-field approximation~\cite{SM}.
From the master equation, we can obtain the time evolution equation for the local density at a site $i$ with a spin $s$, $\rho_{i, s} (t)$, by neglecting the microscopic fluctuation and correlation~\cite{Solon2013,Solon2015a} as
\begin{align}
\frac{\partial \rho_{i, s}}{\partial t} = & \sum_{l = \hat{x}, \hat{y}, -\hat{x}, -\hat{y}} J (1 + \varepsilon \delta_{s, l}) [\rho_{i - l, s} (1 - \rho_i) w^\mathrm{MF}_{i - l \to i} \nonumber \\
& - \rho_{i, s} (1 - \rho_{i + l}) w^\mathrm{MF}_{i \to i + l}] - h (4 \rho_{i, s} - \rho_i),
\end{align}
where, $\rho_i := \sum_s \rho_{i, s}$ and $w^\mathrm{MF}_{i \to j}$ is the mean-field version of $w_{i \to j}$~\cite{SM}.
Focusing on the moderate spatial variation of $\rho_{i, s}$ with respect to the lattice constant $a$, we may replace $\rho_{i, s}$ by $\rho_s (\bm{x})$ and expand $\rho_{i + l, s}$ as $\rho_{i + l, s} \simeq [1 + a l \cdot \nabla + (a l \cdot \nabla)^2 / 2] \rho_s (\bm{x})$.
In the same spirit, we may expand $w^\mathrm{MF}_{i \to i + l}$ as $w^\mathrm{MF}_{i \to i + l} \simeq 1 - 2 U a l \cdot \nabla \rho (\bm{x}) - U (a l \cdot \nabla)^2 \rho (\bm{x})$.
Further, we focus on the temporally slow mode, i.e., the density field $\rho (\bm{x}, t) := \sum_s \rho_s (\bm{x}, t)$, which is important around the critical point, and use the adiabatic approximation~\cite{Speck2014,Speck2015}.
Finally, we obtain the equation for $\rho (\bm{x}, t)$ as
\begin{equation}
\frac{\partial \rho}{\partial t} = \nabla \cdot M (\rho) \nabla \frac{\delta \mathcal{F}_\mathrm{eff}}{\delta \rho}.
\label{Eq:rhoeq}
\end{equation}
Here, $M (\rho)$ $:= (1 + \varepsilon / 4) J a^2 (1 - \rho) \rho$ represents the mobility, and $\mathcal{F}_\mathrm{eff}$ $:= \int \mathrm{d} \bm{x} f (\rho)$ denotes the effective free energy, with
\begin{align}
f(\rho) = & \left[ 1 + \frac{\varepsilon^2 J}{(8 + 2 \varepsilon) h} \right] \rho \ln \rho + (1 - \rho) \ln (1 - \rho) \nonumber \\
& + 2 \left[ U  - \frac{\varepsilon^2 J }{(16 + 4 \varepsilon) h} \right] \rho^2,
\label{Eq:FreeEnergyDensity}
\end{align}
showing that the activity simply works as an additional attractive interaction~\cite{Speck2014,Speck2015,Farage2015,Fodor2016}, as well as breaking the particle-hole symmetry in the entropic terms.

To investigate the mean-field critical point, we expand $f (\rho)$ with respect to $\phi (\bm{x}, t)$ [$:= \rho (\bm{x}, t) - \overline{\rho}$] as $f = A_2 \phi^2 + A_3 \phi^3 + A_4 \phi^4 + O (\phi^5)$, where we omit the $O (\phi^0, \phi^1)$ terms since they do not contribute to Eq.~\eqref{Eq:rhoeq}.
The spinodal line is obtained by
\begin{equation}
A_2 = 1 + 4 U (1 - \overline{\rho}) \overline{\rho} + \frac{\varepsilon^2 J}{(8 + 2 \varepsilon) h} (1 - \overline{\rho}) (1 - 2 \overline{\rho}) =0,
\label{Eq:A2}
\end{equation}
and the critical point by further restraining
\begin{equation}
A_3 = 2 \overline{\rho} - 1- \frac{\varepsilon^2 J}{(8 + 2 \varepsilon) h} (1 - \overline{\rho})^2=0.
\label{Eq:A3}
\end{equation}

Based on Eqs.~\eqref{Eq:A2} and \eqref{Eq:A3}, we obtain the mean-field critical points and spinodal lines in the $\overline{\rho}$-$U$ plane [Fig.~\ref{Fig:PhaseDiagram}(c)] and in the $\overline{\rho}$-$\varepsilon$ plane [Fig.~\ref{Fig:PhaseDiagram}(d)].
The critical points form a line in the $\overline{\rho}$-$U$-$\varepsilon$ space in a similar way to those observed in the simulation [Figs.~\ref{Fig:PhaseDiagram}(a) and (b)], which suggests that the mean-field approximation captures the qualitative behavior of the critical line.
To clearly show the connection between the EPS and MIPS critical points within the mean-field approximation, we obtain the critical line in the $U$-$\varepsilon$ plane [Fig.~\ref{Fig:PhaseDiagram}(e)] by making $\overline{\rho}$ depend on $\varepsilon$ so that Eq.~\eqref{Eq:A3} is satisfied.
Here, the intersection of the critical line and $\varepsilon = 0$ represents the EPS critical point, and part of the critical line for $U \geq 0$ corresponds to the MIPS critical points.

\textit{Universality of the critical line.}
By using a modified version of the recently proposed sub-box method~\cite{Siebert2018,Partridge2019,Maggi2020}, we calculate the critical exponents of the critical line, especially for two cases with both activity and attractive interaction: varying $\varepsilon$ with $U = -1$ and varying $U$ with $\varepsilon = 0.1$.
For both of these cases, the critical density $\rho_\mathrm{c}$ is around $0.5$ based on Figs.~\ref{Fig:PhaseDiagram}(a) and (b), and we set $\rho = 0.5$ in the following.
By considering rectangular systems with the size $10 L \times L$, we take the steady-state configurations from four sub-boxes with the size $L \times L$ [Fig.~\ref{Fig:FSS}(a)], and $\braket{\cdots}$ represents the average over all the independent samples and sub-boxes.

We first focus on varying $\varepsilon$ with $U = -1$.
Defining $\Delta \rho_L := \rho_L - \overline{\rho}$, where $\rho_L$ is the density in the sub-box, we first calculate the Binder ratio $Q_L$ [$:= \braket{(\Delta \rho_L)^2}^2 / \braket{(\Delta \rho_L)^4}$] [Fig.~\ref{Fig:FSS}(b)].
From the approximate intersection of $Q_L (\varepsilon)$ curves in Fig.~\ref{Fig:FSS}(b), we estimate the critical point~\cite{Siebert2018} as $\varepsilon_\mathrm{c} \simeq 1.07$.
At $\varepsilon = \varepsilon_\mathrm{c}$, according to the scaling hypothesis~\cite{Siebert2018}, we can obtain $\partial Q_L / \partial \varepsilon \sim L^{1 / \nu}$, $\chi_L := \braket{(N_L - \braket{N_L})^2} / \braket{N_L} \sim L^{\gamma / \nu}$, and $\braket{(\Delta \rho_L)^2} \sim L^{-\beta / \nu}$, where $N_L$ is the particle number in the sub-box, and $\nu$, $\beta$, and $\gamma$ are the critical exponents.
Comparing these size scalings with the numerical data [Fig.~\ref{Fig:FSS}(d)], we find that the critical exponents at $(U_\mathrm{c}, \varepsilon_\mathrm{c}) \simeq (-1, 1.07)$ are consistent with the 2D Ising universality ($\nu = 1$, $\beta = 1 / 8$, and $\gamma = 7 / 4$).

The corresponding results for varying $U$ with $\varepsilon = 0.1$ are shown in Figs.~\ref{Fig:FSS}(c) and (e), from which we find that the critical exponents at $(U_\mathrm{c}, \varepsilon_\mathrm{c}) \simeq (-1.76, 0.1)$ are also consistent with the Ising universality.
Further, as is well known~\cite{Siebert2018}, the EPS critical point with $\varepsilon = 0$ belongs to the Ising universality class (see \cite{SM} for confirmation in our model).
Lastly, also for the MIPS critical point with $U = 0$, the obtained size scalings seem consistent with the Ising universality~\cite{SM}, as observed in similar active lattice gas models~\cite{Partridge2019} and in Active Ornstein-Uhlenbeck particles~\cite{Maggi2020}.
These results imply that the whole critical line, which connects the EPS and MIPS critical points, belongs to the 2D Ising universality class.

For the critical points obtained above, we examine the dynamical scaling of the domain size $R (t) \sim t^{1/z}$ after a quench from a random configuration in a square system with $\overline{\rho} = 0.5$, where $z$ is the dynamical critical exponent.
Here we define $R (t)$ as the first zero of $C_\mathrm{a} (r, t)$ \{$:= [C (r \hat{x}, t) + C (r \hat{y}, t)] / 2$\}, where $C (\bm{r}, t)$ $[:= L^{-2} \sum_{\bm{r}_0} \braket{\rho (\bm{r} + \bm{r}_0, t) \rho (\bm{r}_0, t)} - \overline{\rho}^2$] is the density correlation function and $\braket{\cdots}$ represents the average over all the independent samples.
The time evolution of $R (t)$ at both $(U, \varepsilon) = (-1, 1.07)$ [Fig.~\ref{Fig:Dynamics}(a)] and $(-1.76, 0.1)$ [Fig.~\ref{Fig:Dynamics}(b)] is consistent with $z = 15 / 4$, the exponent for the 2D spin-exchange Ising universality~\cite{Alexander1994}, as observed in active lattice gas models~\cite{Partridge2019}.

We also perform a deep quench to the phase-separated regime with $(U, \varepsilon) = (-1, 2)$ [Fig.~\ref{Fig:Dynamics}(c)] and find that $R (t)$ shows the Lifshitz–Slyozov–Wagner (LSW) law ($z = 3$)~\cite{Lifshitz1961,Wagner1961}, which has been known to appear in EPS~\cite{Bray1994} and also observed in active lattice gas models~\cite{Thompson2011,Partridge2019}.
The LSW law holds even when the configuration is anisotropic [Fig.~\ref{Fig:Dynamics}(c)], as also demonstrated in equilibrium~\cite{Zhdanov2001} and driven~\cite{Hurtado2002} anisotropic lattice gas models.

\textit{Discussion and conclusions.}
In this Letter, we have studied the lattice gas model with activity and nearest-neighbor interaction.
By MC simulations, we have found that the MIPS and EPS critical points are connected by a critical line, which we can qualitatively reproduce within the mean-field approximation.
We have also investigated both the static and dynamical critical exponents for the critical line by the finite-size scaling analysis, and found that the whole critical line belongs to the 2D spin-exchange Ising universality class.
Further, we confirmed that the LSW law appears for a deep quench toward both attractive interaction and activity.

Our results suggest that activity-induced violation of detailed balance is inessential for the critical phenomena in the motility-induced phase separation; the activity $\varepsilon$ only enters as a parameter in the mean-field free energy [Eq.~\eqref{Eq:FreeEnergyDensity}], which is consistent with the RG analysis of the Active Model B+~\cite{Caballero2018}.
This picture is consistent with the observed LSW law, which reflects the process of reducing the interface free energy between the high-density and low-density phases in the case of EPS~\cite{Bray1994}.
Recently, intracellular phase separation of proteins/mRNAs has been observed, and the functions and mechanism of the liquid droplet formation have been discussed~\cite{Brangwynne2009,Banani2017,Shin2017}.
Our result clarifies that the MIPS and EPS are indistinguishable at the macro-scale observed in common cell experiments, indicating the potential role of activity, fueled for instance by enzyme catalysis~\cite{Jee2017,Jee2018} in the liquid droplet formation in cells.



\textit{Acknowledgments.}
We are grateful to Hiroyoshi Nakano and Michio Tateno for fruitful discussions.
We are also thankful to Takaki Yamamoto for helpful comments.
K.A. is supported by JSPS KAKENHI Grant No. JP20K14435, and the Interdisciplinary Theoretical and Mathematical Sciences Program (iTHEMS) at RIKEN.
K.K is supported by JSPS KAKENHI Grants No. JP18H04760, No. JP18K13515, No. JP19H05275, No. JP19H05795, and by Research Grant from HFSP (Ref.-No: RGY0081/2019).
The numerical calculations have been performed on cluster computers at RIKEN iTHEMS.

\newpage

\onecolumngrid

\renewcommand{\thepage}{S\arabic{page}}  
\renewcommand{\thesection}{S\arabic{section}}   
\renewcommand{\thefigure}{S\arabic{figure}}
\renewcommand{\theequation}{S\arabic{equation}}
\setcounter{page}{1}
\setcounter{section}{0}
\setcounter{figure}{0}
\setcounter{equation}{0}

\begin{center}
\textbf{ \large Supplemental Material for \\ Universality of active and passive phase separation in a lattice model}\\ [.1cm]
{Kyosuke Adachi and Kyogo Kawaguchi}\\ [.1cm]
{(Dated: \today)}\\
\end{center}

\section{Simulation of the lattice gas model}

\subsection{Simulation procedure}

By discretizing time, we perform Monte Carlo (MC) simulations corresponding to the lattice gas model [Fig.~1(a) in the main text].
In this model, each particle with a spin $s$ ($= \hat{x}$, $\hat{y}$, $-\hat{x}$, or $-\hat{y}$) can stochastically (i) hop to a nearest-neighbor site if empty or (ii) flip the spin with a rate $h$, where $\hat{a}$ is the unit translation parallel to the $a$\nobreakdash-axis.
For hopping from site $i$ to an adjacent site $j$, we set a higher rate $(1 + \varepsilon) w_{i \to j} J$ if the hopping is in the same direction as the spin, and a lower rate $w_{i \to j} J$ otherwise, using the activity parameter $\varepsilon$ ($\geq 0$).
We set $w_{i \to j} = 1 - \tanh (\Delta E_{i \to j} / 2)$ with $k_\mathrm{B} T = 1$, where $\Delta E_{i \to j}$ is the increase in the total interaction energy due to hopping, with the nearest-neighbor interaction energy $U$ (repulsive for $U > 0$ and attractive for $U < 0$).
In all the simulations, we set $h / J = 0.01$.

First, we randomly choose a particle, say, at site $i$ with spin $s$.
Then, we randomly choose a direction from $\{ \hat{x}, \hat{y}, -\hat{x}, -\hat{y}\} \setminus \{ s \}$ and update $s$ to the chosen direction with a probability $3 h / 8 J (1 + \varepsilon)$.
Lastly, we randomly choose a direction (we call $l$) from $\{ \hat{x}, \hat{y}, -\hat{x}, -\hat{y}\}$ and move the particle to the adjacent site $i + l$ if empty with a probability $w_{i \to i + l} / 2$ or $w_{i \to i + l} / 2 (1 + \varepsilon)$ for $l = s$ or $l \neq s$, respectively.
We repeat this procedure $N$ (the total particle number) times as 1 MC step.
Note that each flipping/hopping probability is smaller than 1 since $0 < w_{i \to j} < 2$.

\subsection{Finite-size scaling analysis for EPS and MIPS}

\begin{figure*}[t]
\centering
\includegraphics[scale=0.8]{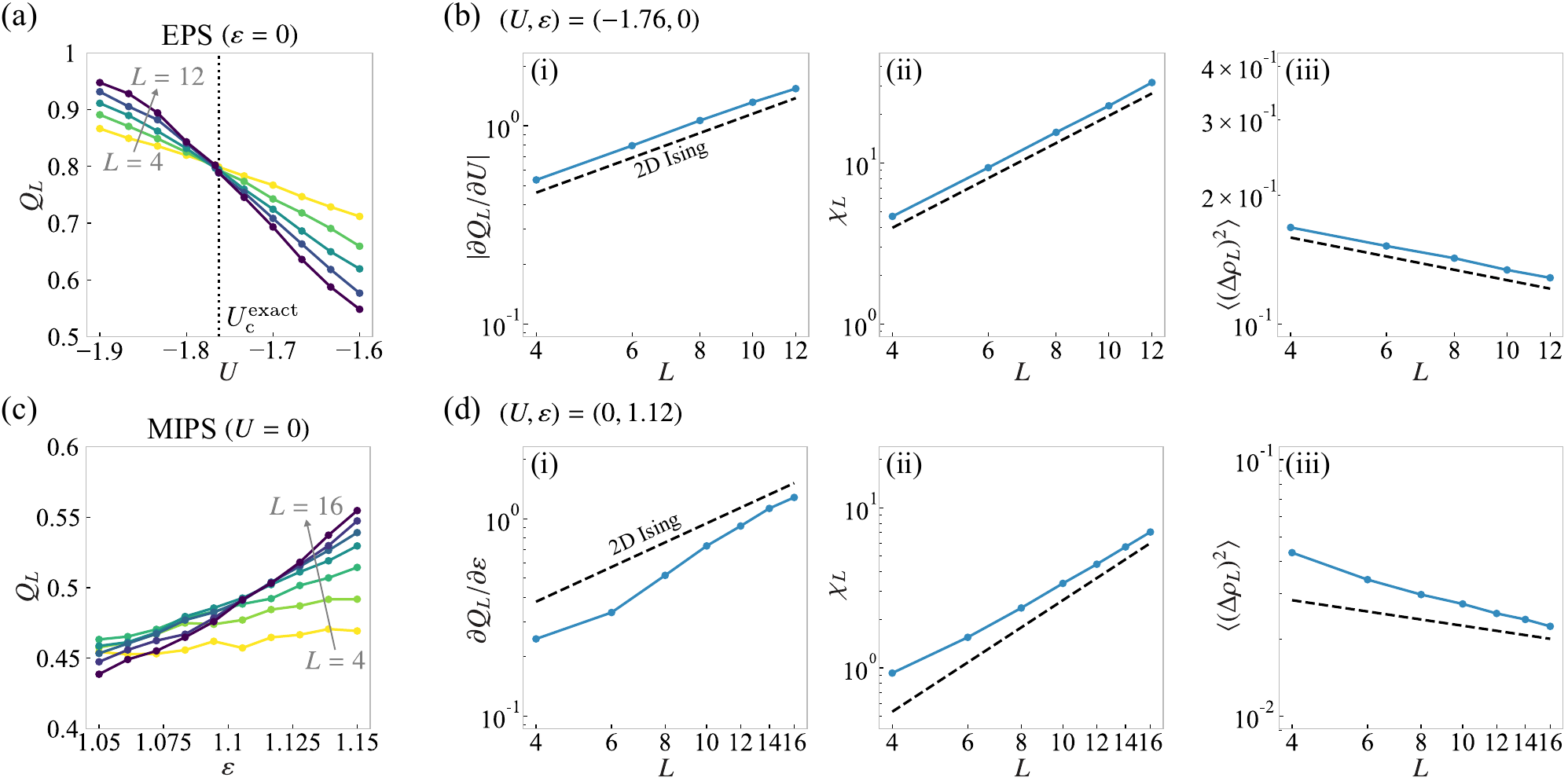}
\caption{(a) The Binder ratio as a function of $U$ with $L = 4, 6, 8, 10, 12$ for EPS ($\varepsilon = 0$).
The black dotted line shows the exact critical point $U_\mathrm{c}^\mathrm{exact} = 2 \ln (1 + \sqrt{2})$
(b) Size scalings of (i) derivative of the Binder ratio ($\sim L^{1/\nu}$), (ii) susceptibility ($\sim L^{\gamma/\nu}$), and (iii) density fluctuation ($\sim L^{-\beta/\nu}$) for EPS ($\varepsilon = 0$).
For comparison, we show the size scalings for the 2D Ising model ($\nu = 1$, $\beta = 1 / 8$, and $\gamma = 7 / 4$; black dashed line).
(c) The Binder ratio as a function of $\varepsilon$ with $L = 4, 6, 8, 10, 12, 14, 16$ for MIPS ($U = 0$).
(d) Size scalings [counterparts of (b)] for MIPS ($U = 0$).
For (a, b) [(c, d)], we performed 480 independent simulations and sampled 41 (31) configurations at intervals of $10^5$ MC steps after $8 \times 10^6$ ($10^6$) MC steps in each simulation  with the fully phase-separated (random) initial configurations.
We used $\overline{\rho} = 0.5$ and $\overline{\rho} = 0.55$ for (a, b) and (c, d), respectively, and $h / J = 0.01$ for all figures.}
\label{FigS:FSS}
\end{figure*}

We show the results of the finite-size scaling analysis for EPS with $\varepsilon = 0$ [Figs.~\ref{FigS:FSS}(a) and (b)] and MIPS with $U = 0$ [Figs.~\ref{FigS:FSS}(c) and (d)].
For EPS, we set $\rho = 0.5$, and the crossing of the Binder ratio $Q_L$ [Fig.~\ref{FigS:FSS}(a)] shows the critical point $(U_\mathrm{c}, \varepsilon_\mathrm{c}) \simeq (-1.76, 0)$.
The obtained $U_\mathrm{c}$ is close to the exact value~\cite{Onsager1944}, $U_\mathrm{c}^\mathrm{exact} = 2 \ln (1 + \sqrt{2}) = 1.7627...$, which suggests that the sub-box method [Fig.~3(a) in the main text] is working.
As expected, the critical exponents are consistent with the 2D Ising universality [Fig.~\ref{FigS:FSS}(b)].
For MIPS with $U = 0$, we set $\rho = 0.55$ considering the shift of $\rho_\mathrm{c}$ [Fig.2(b) in the main text].
Based on the crossing of the Binder ratio $Q_L$ for $L \geq 10$ [Fig.~\ref{FigS:FSS}(c)], we estimate the critical point as $(U_\mathrm{c}, \varepsilon_\mathrm{c}) \simeq (0, 1.12)$, although the crossing is not as clear as the cases with negative $U$.
The size scalings seem consistent with the 2D Ising universality [Fig.~\ref{FigS:FSS}(d)], though we do not reach the scaling regime due to the limited system size.

\begin{figure*}[t]
\centering
\includegraphics[scale=0.8]{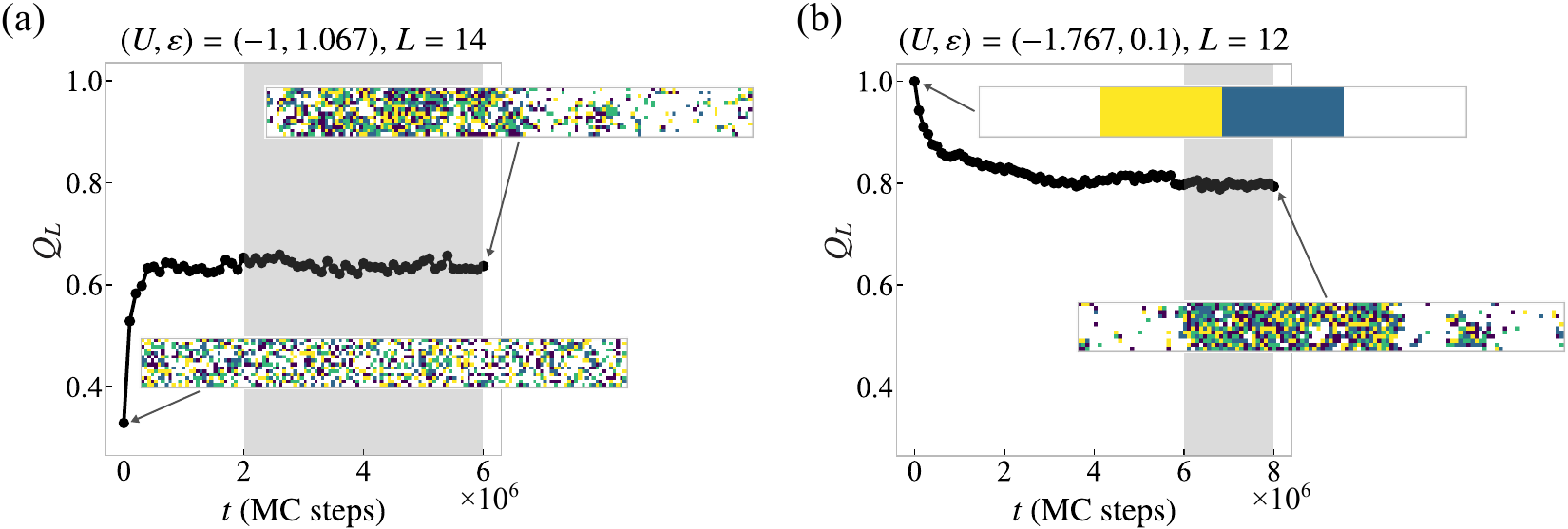}
\caption{Time evolution of the Binder cumulant for two parameter sets around the critical line: (a) $(U, \varepsilon) = (-1, 1.067)$ with $L = 14$ and (b) $(U, \varepsilon) = (-1.767, 0.1)$ with $L = 12$, which correspond to the cases in Figs.~3(b) and (c), respectively.
We also show typical final configurations as well as the (a) random or (b) fully phase-separated initial configurations, where the yellow, green, blue, and purple dots represent the particles with $s = \hat{x}$, $\hat{y}$, $-\hat{x}$, and $-\hat{y}$, respectively.
In the finite-size scaling analysis, we used configurations at time points shown as the gray region.}
\label{FigS:conv}
\end{figure*}

\begin{figure*}[t]
\centering
\includegraphics[scale=0.8]{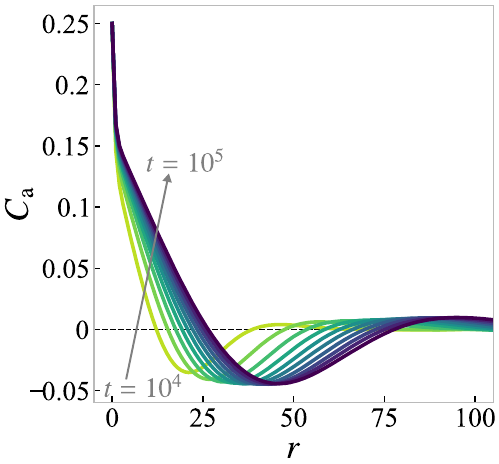}
\caption{Time evolution of the correlation function averaged in the axial direction, $C_\mathrm{a} (r, t)$, between $10^4$ and $10^5$ MC steps for $(U, \varepsilon) = (-1, 2)$, which corresponds to Fig.~4(c) in the main text.
The first zero of $C_\mathrm{a} (r, t)$ at each time $t$ represents the domain size $R(t)$.
Note that $C_\mathrm{a} (0, t) = \overline{\rho} (1 - \overline{\rho})$ and thus $C_\mathrm{a} (0, t) = 0.25$ for $\overline{\rho} = 0.5$.}
\label{FigS:corr}
\end{figure*}

\subsection{Relaxation dynamics}

In the finite-size scaling analysis, we sample configurations of the steady state, which is realized after relaxation from the initial configuration.
In Fig.~\ref{FigS:conv}, we show typical time evolution of the Binder cumulant for two kinds of parameter sets around the critical line: (a) $(U, \varepsilon) = (-1, 1.067)$ and (b) $(U, \varepsilon) = (-1.767, 0.1)$.
For $U = -1$, the dynamics of $Q_L$ shows the relaxation to the steady state from the random configuration [Fig.~\ref{FigS:conv}(a)].
For $\varepsilon = 0.1$, we perform simulations from the fully phase-separated configuration to accelerate the relaxation for negatively large $U$, and the dynamics of $Q_L$ represents the relaxation process [Fig.~\ref{FigS:conv}(b)].
Similarly, we use the fully phase-separated initial configuration in simulations for Figs.~3(c) and (e) in the main text and Figs.~\ref{FigS:FSS}(a) and (b).

The domain size $R(t)$ is determined by the first zero of the correlation function $C_\mathrm{a} (r, t)$ defined in the main text.
Figure~\ref{FigS:corr} is an example of the time dependence of $C_\mathrm{a}$ for the parameters corresponding to Fig.~4(c) in the main text, and we see the growth of $R(t)$ as time passes.

\section{Mean-field approximation}

We explain the details of the mean-field approximation used in the main text.
In the following, we use $\braket{\cdots}_t$ as the average with respect to the probability $P (\{ n_{i, s} \}, t)$ for the configuration $\{ n_{i, s} \}$ at time $t$, where $n_{i, s}$ ($= 0$ or 1) is the local occupancy.
Based on the master equation, which describes the time evolution of $P (\{ n_{i, s} \}, t)$, we can obtain the equation for $\braket{n_{i,s}}_t$ as
\begin{equation}
\frac{\partial \braket{n_{i,s}}_t}{\partial t} = \sum_{l = \hat{x}, \hat{y}, -\hat{x}, -\hat{y}} J (1 + \varepsilon \delta_{s, l}) [\braket{n_{i - l, s} (1 - n_i) w_{i - l \to i}}_t - \braket{n_{i, s} (1 - n_{i + l}) w_{i \to i + l}}_t] - h (4 \braket{n_{i, s}}_t - \braket{n_i}_t).
\end{equation}
Here, $n_i := \sum_s n_{i, s}$.
We neglect the second and higher-order correlations within the mean-field approximation~\cite{Solon2013,Solon2015a}, which leads to Eq.~(1) in the main text:
\begin{equation}
\frac{\partial \rho_{i, s}}{\partial t} = \sum_{l = \hat{x}, \hat{y}, -\hat{x}, -\hat{y}} J (1 + \varepsilon \delta_{s, l}) [\rho_{i - l, s} (1 - \rho_i) w^\mathrm{MF}_{i - l \to i} - \rho_{i, s} (1 - \rho_{i + l}) w^\mathrm{MF}_{i \to i + l}] - h (4 \rho_{i, s} - \rho_i),
\label{EqS:rhoeqlat}
\end{equation}
where $\rho_{i,s} (t) := \braket{n_{i,s}}_t$, $\rho_{i} (t) := \sum_s \rho_{i, s} (t)$, and $w^\mathrm{MF}_{i \to j} = 1 - \tanh [(E^\mathrm{MF}_j - E^\mathrm{MF}_i) / 2]$ with $E^\mathrm{MF}_i := \sum_{l = \hat{x}, \hat{y}, -\hat{x}, -\hat{y}} U \rho_{i + l}$.

Focusing on the moderate spatial variation of $\rho_{i, s}$ with respect to the lattice constant $a$, we replace $\rho_{i, s}$ by $\rho_s (\bm{x})$ and $\rho_{i}$ by $\rho (\bm{x})$, expand $\rho_{i + l, s}$ as $\rho_{i + l, s} \simeq [1 + a l \cdot \nabla + (a l \cdot \nabla)^2 / 2] \rho_s (\bm{x})$, and expand $w^\mathrm{MF}_{i \to i + l}$ as $w^\mathrm{MF}_{i \to i + l} \simeq 1 - 2 U a l \cdot \nabla \rho (\bm{x}) - U (a l \cdot \nabla)^2 \rho (\bm{x})$.
Substituting these expressions in Eq.~\eqref{EqS:rhoeqlat}, we obtain, up to $O(a^2)$,
\begin{align}
\frac{\partial \rho_s}{\partial t} = & J a^2 \{ (1 - \rho) \nabla^2 \rho_s + \rho_s \nabla^2 \rho + 4U \nabla \cdot [(1 - \rho) \rho_s \nabla \rho] \} + \frac{\varepsilon J a^2}{2} \{ (1 - \rho) (s \cdot \nabla)^2 \rho_s + \rho_s (s \cdot \nabla)^2 \rho + 4 U (s \cdot \nabla) [(1 - \rho) \rho_s (s \cdot \nabla \rho)] \} \nonumber \\
& - \varepsilon J a (s \cdot \nabla) [(1 - \rho) \rho_s] - h (4 \rho_s - \rho).
\label{EqS:rhoeqcont}
\end{align}

To obtain the critical point and spinodal line, which are determined by the temporally slow mode, i.e., the density field $\rho (\bm{x}, t)$ in our model, we use the adiabatic approximation~\cite{Speck2014,Speck2015}.
First, using the ``magnetization field'' $m_\alpha := \rho_{\hat{\alpha}} - \rho_{-\hat{\alpha}}$ ($\alpha = x$ or $y$) and ``nematicity field'' $\nu := \rho_{\hat{x}} + \rho_{-\hat{x}} - \rho_{\hat{y}} - \rho_{-\hat{y}}$ in addition to the density field $\rho = \sum_s \rho_s$, we rewrite Eq.~\eqref{EqS:rhoeqcont} as
\begin{align}
\frac{\partial \rho}{\partial t} = & J a^2 \{ \nabla^2 \rho + 4U \nabla \cdot [(1 - \rho) \rho \nabla \rho] \} + \frac{\varepsilon J a^2}{4} \{ \nabla^2 \rho + (1 - \rho)({\partial_x}^2 - {\partial_y}^2) \nu + \nu ({\partial_x}^2 - {\partial_y}^2) \rho \nonumber \\
& + 4 U \nabla \cdot [(1 - \rho) \rho \nabla \rho] + 4 U \partial_x [(1 - \rho) \nu \partial_x \rho] - 4 U \partial_y [(1 - \rho) \nu \partial_y \rho] \} - \varepsilon J a \nabla \cdot \bm{m},
\label{EqS:rhoeq} \\
\frac{\partial m_\alpha}{\partial t} = & J a^2 \{ (1 - \rho) \nabla^2 m_\alpha + m_\alpha \nabla^2 \rho + 4 U \partial_\alpha [(1 - \rho) m_\alpha \partial_\alpha \rho] \} + \frac{\varepsilon J a^2}{2} \{ (1 - \rho) {\partial_\alpha}^2 m_\alpha + m_\alpha {\partial_\alpha}^2 \rho + 4 U \partial_\alpha [(1 - \rho) m_\alpha \partial_\alpha \rho] \} \nonumber \\
& - \frac{1}{2} \varepsilon J a \partial_\alpha \{ (1 - \rho) [\rho + \hat{\alpha} \cdot (\hat{x} - \hat{y}) \nu] \} - 4 h m_\alpha,
\label{EqS:meq} \\
\frac{\partial \nu}{\partial t} = & J a^2 \{ (1 - \rho) \nabla^2 \nu + \nu \nabla^2 \rho + 4 U \nabla \cdot [(1 - \rho) \nu \nabla \rho] \} + \frac{\varepsilon J a^2}{4} \{ ({\partial_x}^2 - {\partial_y}^2) \rho + (1 - \rho) \nabla^2 \nu + \nu \nabla^2 \rho \nonumber \\
& + 4 U \partial_x [(1 - \rho) \rho \partial_x \rho] - 4 U \partial_y [(1 - \rho) \rho \partial_y \rho] + 4 U \nabla \cdot [(1 - \rho) \nu \nabla \rho] \} - \varepsilon J a \{ \partial_x [(1 - \rho) m_x] - \partial_y [(1 - \rho) m_y] \} - 4 h \nu.
\label{EqS:nueq}
\end{align}
Assuming the spatially slow variation of $\rho$, $\bm{m}$, and $\nu$, we see from Eqs.~\eqref{EqS:meq} and \eqref{EqS:nueq} that $\bm{m}$ and $\nu$ will be rapidly relaxed with a timescale $\sim 1 / 4 h$.
Thus, focusing on the relaxation timescale of $\rho$ ($\gg 1 / 4 h$), we can approximately set $\partial_t m_\alpha = 0$ and $\partial_t \nu = 0$ in Eqs.~\eqref{EqS:meq} and \eqref{EqS:nueq}, respectively.
Within this adiabatic approximation, we can show $\bm{m} = - \varepsilon J a \nabla [(1 - \rho) \rho] / 8 h + O (a^2)$ and $\nu = O (a^2)$, thereby obtaining
\begin{equation}
\frac{\partial \rho}{\partial t} = \left( 1 + \frac{\varepsilon}{4} \right) J a^2 \{ \nabla^2 \rho + 4 U \nabla \cdot [(1 - \rho) \rho \nabla \rho] \} + \frac{\varepsilon^2 J^2 a^2}{8 h} \nabla \cdot \{ (1 - \rho) \nabla [(1 - \rho) \rho] \},
\label{EqS:rhoeqad}
\end{equation}
up to $O (a^2)$.
Using the mobility $M (\rho) := (1 + \varepsilon / 4) J a^2 (1 - \rho) \rho$ and the effective free energy $\mathcal{F}_\mathrm{eff} := \int \mathrm{d} \bm{x} f (\rho)$, where
\begin{equation}
f(\rho) = \left[ 1 + \frac{\varepsilon^2 J}{(8 + 2 \varepsilon) h} \right] \rho \ln \rho + (1 - \rho) \ln (1 - \rho) + 2 \left[ U  - \frac{\varepsilon^2 J }{(16 + 4 \varepsilon) h} \right] \rho^2,
\end{equation}
we can rewrite Eq.~\eqref{EqS:rhoeqad} as Eq.~(2) in the main text, i.e.,
\begin{equation}
\frac{\partial \rho}{\partial t} = \nabla \cdot M (\rho) \nabla \frac{\delta \mathcal{F}_\mathrm{eff}}{\delta \rho}.
\end{equation}

%

\end{document}